\documentclass[twocolumn]{autart}    

\usepackage{graphicx}          

\usepackage{amsfonts}
\usepackage{amsmath}
\usepackage{cite}
\usepackage{color}
\usepackage{txfonts}

\usepackage{mdwlist}
\usepackage{enumitem}

\allowdisplaybreaks[4]

\newtheorem{theorem}{Theorem}
\newtheorem{lemma}{Lemma}

\newtheorem{remark}{Remark}
\newtheorem{proposition}{Proposition}

\newtheorem{example}{Example}
\newtheorem{definition}{Definition}

\newcommand{\R}{\mathbb{R}}

\newcommand{\cl}{\text{cl}}

\usepackage{xcolor}

\begin{document}

\begin{frontmatter}

\title{Inverse optimal design of input-to-state stabilizing homogeneous controllers for nonlinear homogeneous systems} 

\thanks[footnoteinfo]{This paper was not presented at any IFAC meeting. This work was supported by the MOE Tier 2 Grant under the Project No. MOE-000488-00 from the the Ministry of Education of Singapore, and was supported by Natural Science Foundation of Jiangsu Province under Grant BK20230261, and was supported by Jiangsu Funding Program for Excellent Postdoctoral Talent under Grant: 2025ZB179. Corresponding author H.~Yu.}

\author[nus,nusri]{Kaixin Lu}\ead{lukaixin@nus.edu.sg},
\author[ntu,tongji]{Ziliang Lyu}\ead{ziliang.lyu@ntu.edu.sg},
\author[nus,nusri]{Haoyong Yu}\ead{bieyhy@nus.edu.sg}

\address[nus]{Department of Biomedical Engineering, National University of Singapore, Singapore.}
\address[nusri]{National University of Singapore (Suzhou) Research Institute, Suzhou, Jiangsu, China.}
\address[ntu]{School of Electrical and Electronic Engineering, Nanyang Technological University, Singapore.}
\address[tongji]{Department of Control Science and Engineering, Tongji University, Shanghai, China.}


\begin{abstract}
    This work studies the inverse optimality of input-to-state stabilizing controllers with input-output stability guarantees for nonlinear homogeneous systems. We formulate a new inverse optimal control problem, where the cost functional incorporates penalties on the output, in addition to the state, control and disturbance as in current related works. One benefit of penalizing the output is that the resulting inverse optimal controllers can ensure both input-to-state stability and input-output stability. We propose a technique for constructing the corresponding meaningful cost functional by using homogeneity properties, and provide sufficient conditions on solving the inverse optimal gain assignment problem. We show that homogeneous stabilizability of homogeneous systems in the case without disturbance is sufficient for the solvability of inverse optimal gain assignment problem for homogeneous systems.
\end{abstract}

\end{frontmatter}

\section{Introduction}

Optimality guarantees many desirable properties for closed-loop systems. In direct optimal control, constructing an optimal control law for nonlinear systems poses considerable challenges, primarily due to the difficulty of solving a Hamilton-Jacobi-Isaacs (HJI) equation. This motivates the study of inverse optimal control. Inverse optimal control was pioneered by Kalman \cite{kalman1964linear} and has received substantial attention over the past decades \cite{freeman1996inverse,Nakamura2009tac,krstic1998inverse,deng1997stochastic,
lu2023auto,li1997optimal,Belhadjoudja_2024_ECC,Anderson_1990,Moylan_1973_tac,
Sepulchre_1997}. Inverse optimal control approaches avoid the task of solving an HJI equation. Moreover, it is well-known that inverse optimal controllers possess gain margin with respect to uncertainties in the control gain.

Input-to-state stability (ISS) provides a tool to measure the influence of the initial conditions and the magnitude of disturbance inputs on the stability of internal states \cite{Sontag1995scl}. In \cite{krstic1998inverse}, Krstic and Li proposed a promising inverse optimal ISS control framework, and showed that input-to-state stabilizability is equivalent to the solvability of a two-player differential game problem. Compared with pure ISS controllers, inverse optimal ISS controllers possess the advantage of ensuring ISS for closed-loop systems even though there are uncertainties in the control gain. Over the past two decades, the methods in \cite{krstic1998inverse} have been widely used to reveal the connections between inverse optimality and other properties of nonlinear systems, including asymptotic stabilization in probability \cite{deng1997stochastic}, adaptive stabilization \cite{li1997optimal}, input-to-state practical stability \cite{lu2023auto}, and set invariance \cite{Krstic_inverse_safety_2024}.

Note that, for practical control problems, after ISS is guaranteed, it is often desirable to understand how the disturbance inputs affect a certain closed-loop signal that the control designer is interested in, so as to achieve a certain level of disturbance attenuation. For example, in nonlinear $H_\infty$ control problem \cite{Isidori_1992,van_der_schaft_tac_1992,Linwei_ijc_1996}, the closed-loop system is required to exhibit input-output behavior in the sense of finite-gain $L_2$ stability. One way to characterize this relation is input-output stability (IOS), where the output can be viewed as a signal that the designer is interested in. Moreover, ISS describes the relationships between the state and the disturbance input from an $L_\infty$ perspective \cite{sontag-integral-ISS}. However, in some control systems, it is challenging to guarantee that an ``output of interest'' belongs to $L_\infty$. An example is the well-known modular adaptive control design based on passive identifiers \cite{krstic_ijc_1994,Krstic_1995}, where the derivative of the parameter estimate is $L_2$, but it is unknown whether this signal belongs to $L_\infty$. Therefore, it is beneficial to consider an ISS controller with an additional IOS guarantee in $L_p$ sense. Motivated by the significance of inverse optimality, it is important to study what kind of optimal criterion that an ISS controller with an IOS guarantee is inverse optimal with respect to. This problem is difficult to address, because, when the output is involved, it is difficult to find a meaningful cost functional, in which the penalties are required to be non-negative (or non-positive, depending on the context) on the whole state space.

Homogeneity is an intrinsic property of an object and has been proved useful, as it can transform the local properties of a system to the whole domain via a suitable scaling  \cite{Kawski1990ctat,hong_2001_tac,Hermes_1991_de,
Rosier1992scl,Ryan_1995,Bernuau_2013}. Homogeneous control has nice applications on nonsmooth finite-time control \cite{hong_2001_tac,Bhat_mcss_2005}, fixed-time safety filter design \cite{Polyakov_2023}, model predictive control \cite{coron_MPC} and optimal control \cite{Polyakov_2020}. In addition, homogeneous technique can be used for continuous feedback designs for systems that cannot be stabilized via smooth feedback \cite{hong2001auto_homogenous,zhao_2022,yu_2023}. In \cite{Andrieu_V_2008,Bernuau_2013,Bernuau_2014}, it was shown that ISS for a nonlinear system can be verified by establishing its asymptotic stability in the unperturbed case under homogeneity conditions, without finding a concrete ISS Lyapunov function as in conventional cases. Comprehensive results on homogeneous systems and nonlinear control were given in \cite{Moulay2008tac,qian_2015,Polyakov_2020}.

The above observations motivate us to leverage homogeneity properties to establish inverse optimality for ISS controllers with IOS guarantees. Our contributions are as follows. We formulate a new inverse optimal control problem, where the cost functional incorporates penalties on the state, control, disturbance and the output. Note that none of the costs in current related works (see for example \cite{krstic1998inverse}) puts penalty on the output. The benefit of penalizing the output is that the resulting inverse optimal controllers ensure both ISS and IOS. Although such inverse optimal controllers provide IOS guarantees, it is very difficult to construct the corresponding meaningful cost functionals, particularly in cases where the output encompasses both the system state and control input. We propose a technique to address this challenge by using homogeneity properties and provide  sufficient conditions for solving the inverse optimal problem (see Theorem 1). We then show how to satisfy these conditions and prove that homogeneous stabilizability of homogeneous systems in the case without disturbance is sufficient for the solvability of inverse optimal gain assignment problem for homogeneous systems (see Theorem 2).

\textbf{Notations.} In this paper, $|\cdot|$ denotes the Euclidean norm for vectors or absolute value for real numbers. For any fixed $p\in[1,\infty)$, we say that $u:\R_+\rightarrow\R^m$ belongs to $L_p^m$ if $u$ is locally integrable and $\int_0^\infty|u(t)|^pdt<\infty$, and denote the $L_p$ norm of $u$ by $\|u\|_{p}=(\int_0^\infty|u(t)|^pdt)^{1/p}$. We say that $u$ belongs to $L_\infty^m$ if $\sup_{t\geq0}|u(t)|<\infty$, and denote the $L_\infty^m$ norm of $u$ by $\|u\|_\infty=\sup_{t\geq0}|u(t)|$. A function $\alpha:\R_+\rightarrow\R_+$ satisfying $\alpha(0)=0$ is said to be of class $K$ if it is continuous and strictly increasing; it is of class $K_\infty$ if, in addition, $\alpha(s)\rightarrow\infty$ as $s\rightarrow\infty$. A function $\beta:\R_{+}\times\R_{+}\rightarrow\R_{+}$ is said to be of class $KL$ if $s\mapsto\beta(s,t)$ is of class $K$ and $t\mapsto\beta(s,t)$ decreases to zero as $t\rightarrow\infty$.
Given a continuous function $V:\R^n\rightarrow\R$, we denote by $L_\mu V(x)$ the Lie derivative of $V$ along the vector field $\mu$ when it exists in $\R^n$, i.e., $L_\mu V(x)=\lim_{t\rightarrow0^+}[V(\phi(t,x))-V(x)]/t$, where $\phi(t,x)$ is the solution of $\dot{x}=\mu(x)$ passing through $x\in\R^n$ at $t=0$. In particular, when $V(x)$ is continuously differentiable, $L_\mu V(x)=\frac{\partial V}{\partial x}\mu(x)$.

\section{Preliminaries}
\subsection{Homogeneous properties}

To begin with, we review the basic notions and properties of homogeneity.

\begin{definition}[Dilation \cite{Kawski1990ctat}]\rm
A dilation $\Delta_\varepsilon^{(r_1,r_2,...,r_n)}$ is defined with coordinates $x=[x_1,x_2,...,x_n]^T\in\R^n$ by assigning a real number $\varepsilon>0$ and $n$ positive real numbers $r_1,r_2,...,r_n$ such that
\begin{align}\label{dilation}
  \Delta_\varepsilon^{(r_1,r_2,...,r_n)}(x_1,x_2,...,x_n)
  =(\varepsilon^{r_1}x_1,\varepsilon^{r_2}x_2,...,\varepsilon^{r_n}x_n).
\end{align}
To simplify notation, we use $\Delta_\varepsilon^r$ to denote $\Delta_\varepsilon^{(r_1,r_2,...,r_n)}$, and $\Delta_\varepsilon^r x$ to represent $\Delta_\varepsilon^{(r_1,r_2,...,r_n)}(x_1,x_2,...,x_n)$ through out this work.
\end{definition}

\begin{definition}[\cite{Kawski1990ctat}]\rm
For any $x\in\R^n$, a scalar function $F(x)$ is homogeneous of degree $l\in \R$ corresponding to the dilation $\Delta_\varepsilon^r$, if it satisfies $F(\Delta_\varepsilon^rx)
  =\varepsilon^l F(x)$ for all $\varepsilon>0$.
A vector field $\varphi(x)=[\varphi_1(x),\varphi_2(x),...,\varphi_n(x)]^T$ is homogeneous of degree $k\in \R$ corresponding to $\Delta_\varepsilon^r$, if, for $i=1,...,n$, it holds that $\varphi_i(\Delta_\varepsilon^rx)
  =\varepsilon^{k+r_i} \varphi_i(x)$ for all $\varepsilon>0$.
\end{definition}

\begin{definition}\rm
A continuous map $\Gamma:\R^n\rightarrow \R$ is said to be a homogenous norm corresponding to $\Delta_\varepsilon^r$, if it is positive definite\footnote{A function $\Gamma(x)$ is positive definite, if $\Gamma(x)=0$ when $x=0$, and $\Gamma(x)>0$ when $x\in\R^n\backslash\{0\}$.} and homogenous of degree one. In this work, the homogenous norm is chosen as $\Gamma(x)=(\sum_{i=1}^{n}|x_i|^{\nu/r_i})^{1/\nu}$, where $\nu$ is a constant such that $\nu>\max\{r_1,\dots,r_n\}$.
\end{definition}

The follow lemma is given in \cite[Theorem 4.1(iii)]{Bhat_mcss_2005}.

\begin{lemma}[\cite{Bhat_mcss_2005}]\label{lem:bhat-homo-cont}\rm
Suppose that $\psi:\R^n\rightarrow\R$ is continuous on $\R^n\backslash\{0\}$ and homogeneous of degree $l$. Then $\psi$ is continuous on the whole domain $\R^n$ if $l>0$.
\end{lemma}

\subsection{Stability and stabilization concepts}

This subsection reviews the stability and stabilization notions used in this paper. Consider a class of homogenous nonlinear systems with respect to the dilation  $\Delta_\varepsilon^r$
\begin{align}\label{sys}
  \dot{x} & = f(x) + G_1(x)u + G_2(x)w, \nonumber \\
  y & = h(x) + du,
\end{align}
where $x\in \R^n$ is the state, $u\in \R$ is the control input, $w\in \R^{\xi}$ is the disturbance input, and $y\in \R^l$ is the output. $f(x)$, $G_1(x)$, $G_2(x)$ and $h(x)$ are continuous on $\R^n$, and continuously differentiable on $\R^n\backslash\{0\}$, with $f(0)=0$ and $h(0)=0$. Moreover, $f(x)=[f_1,f_2,...,f_n]^T$ is a homogeneous vector field of degree $k>-r_0$ with respect to $\Delta_\varepsilon^r$, where $r_0=\min\{r_1,...,r_n\}$. $G_1(x)=[G_{11},G_{12},...,G_{1n}]^T$ and $G_2(x)=[G_{21}^T,G_{22}^T,...,G_{2n}^T]^T$ are homogeneous vector fields of degree $\varsigma=-r_0$ corresponding  to $\Delta_\varepsilon^r$, where $G_{2i}(x)=[G_{2,i1},G_{2,i2},...,G_{2,i\xi}]^T$ for $i=1,...,n$. $h(x)$ is a homogenous function of degree $k+r_0$ corresponding to $\Delta_\varepsilon^r$. $d$ is a constant vector. Analogous to \cite{Isidori_1992,Doyle1989tac,Isidori_1998}, we assume
\begin{align}\label{control_gain}
  h^Td=0, \quad d^Td = \vartheta^2I,
\end{align}
where $\vartheta>0$ is a constant and $I$ is an identity matrix. Herein, $h^Td=0$ implies that $h(x)$ and $du$ are orthogonal, while $d^Td =\vartheta^2I$ implies that the control weight matrix is the identity, with $\vartheta^2$ as the weighting factor.

\begin{definition}[Homogeneous stabilizability\cite{hong2001auto_homogenous}]\rm
The homogeneous control system
\begin{align}\label{sys_wequ_zero}
  \dot{x} & = f(x) + G_1(x)u,
\end{align}
is homogeneously stabilizable, if there exists a control law $u=\alpha(x)$ such that the closed-loop system $\dot{x}=f(x)+G_1(x)\alpha(x)$ is homogeneous of degree $k$ and the equilibrium $x=0$ is asymptotically stable.
\end{definition}

\begin{definition}[Input-to-state stability\cite{Sontag1995scl}]\rm
System
\begin{align}\label{pre_sys_L2}
    \dot{x} = f(x)+G_2(x)w, \quad y=h(x)
\end{align}
is input-to-state stable, if there exist a $KL$-function $\sigma$ and a $K$-function $\gamma$ such that, for any $w\in L_{\infty}^m$ and $x(0)\in \R^n$,
\begin{align}\label{iss_defi}
  |x(t)|\leq \sigma(|x(0)|,t) + \gamma(\|w_t\|_\infty), \quad \forall t>0
\end{align}
where $w_t$ is the truncation of $w$ at time $t$; namely, $w_t(\tau)=w(\tau)$ if $0\leq\tau\leq t$, and $w_t(\tau)=0$ if $\tau\geq t$.
\end{definition}

In this work, IOS is characterized in $L_p$ sense.

\begin{definition}[Finite-gain $L_p$ stability\cite{IOS_book_2009}]\rm
Given $1\leq p<\infty$, system (\ref{pre_sys_L2}) is finite-gain $L_p$ stable if there exist non-negative constants $\kappa_L$ and $c_0$ such that
    \begin{align}\label{L2}
      ||y||_p \leq \kappa_L||w||_p + c_0.
    \end{align}
\end{definition}

\subsection{Legendre-Fenchel transform}

Let $\gamma$ be a $K_\infty$-function whose derivative $\gamma^\prime$ exists and is also of class $K_\infty$. Denote by $\ell_\gamma$ the Legendre-Fenchel transform of $\gamma$, namely,
\begin{align}\label{Legendre-Fenchel transform}
  \ell_\gamma(s)= s(\gamma^\prime)^{-1}(s) - \gamma((\gamma^\prime)^{-1}(s)),
\end{align}
where $(\gamma^\prime)^{-1}(s)$ is the inverse function of $\gamma'(s)=d\gamma(s)/ds$.

\begin{lemma}[\cite{krstic1998inverse}]\label{lem:LF-transform}\rm
The Legendre-Fenchel transform of a class $K_\infty$ function $\gamma$ possesses the following properties: (i) $\ell\ell_\gamma = \gamma$; and (ii) for any vectors $a$ and $b$,
    \begin{align}\label{young_gamma}
      a^Tb\leq\gamma(|a|)+\ell_\gamma(|b|).
    \end{align}
    Moreover, $a^Tb=\gamma(|a|)+\ell_\gamma(|b|)$ if and only if $b=a\gamma^\prime(|a|)/{|a|}$.
\end{lemma}

\section{Inverse optimal gain assignment via homogeneous feedback}

\subsection{Problem formulation}

We formulate our inverse optimal gain assignment problem for system (\ref{sys}) as follows.

\begin{definition}\label{def:inv-opt}\rm
The inverse optimal gain assignment problem for (\ref{sys}) is solvable if there exist positive definite radially unbounded functions $E(x)$ and $l(x)$, matrix-valued functions $R_1(x)=R_1^T(x)>0$ and $R_2(x)=R_2^T(x)>0$, $K_\infty$-function $\gamma_0$ whose derivative $\gamma_0^\prime$ exists and is also of class $K_\infty$, and a control law $u=\alpha^*(x)$, which is continuous away from the origin with $\alpha(0)=0$ and minimizes the cost functional
\begin{align}\label{Ju_de}
  J(u) & = \sup_{w\in \Omega_w}\bigg\{ \lim_{t\rightarrow\infty}\bigg[
  E(x(t))
  + \int_{0}^{t}\bigg(l(x)
  \nonumber \\
  & \quad \quad
   + u^TR_1(x)u
   + y^TR_2(x)y
   -\gamma_0(|w|)
   \bigg)d\tau
   \bigg]\bigg\},
\end{align}
where $\Omega_w$ is the set of all possible $w$.
\end{definition}

\begin{remark}\rm
In the inverse optimal control problem, we need to search for, not only a control law $u=\alpha^*(x)$, but also functions $l(x)$, $R_1(x)$, $R_2(x)$ and $\gamma_0(s)$. In the most relevant work \cite{krstic1998inverse}, the cost functional was constructed as
\begin{align}\label{Ju_krstic}
  J(u) & = \sup_{w\in \Omega_w}\bigg\{ \lim_{t\rightarrow\infty}\bigg[
  E(x(t))
  \nonumber \\
  & \quad \quad
   + \int_{0}^{t}\bigg(\bar{l}(x)
   + u^T\bar{R}_1(x)u
   -\gamma_0(|w|)
   \bigg)d\tau
   \bigg]\bigg\}.
\end{align}
Compared with \cite{krstic1998inverse}, the essential difference of our cost functional (\ref{Ju_de}) lies in the additional penalty $y^TR_2(x)y$ on the output. As a result, our cost functional (\ref{Ju_de}) can ensure both ISS and IOS.
\end{remark}

Below, we use two examples to illustrate the benefits of penalizing the output. First, we show that an inverse optimal controller may fail to guarantee IOS in $L_2$ sense, if the output is not penalized.

\begin{example}\rm
Consider the system
\begin{align}\label{counterexample}
    \dot{x} = x^3 + u + w,
    \quad
     y  = x
\end{align}
with $x_0=x(0)$. Take a Lyapunov function candidate $V(x)=\frac{1}{4}x^4$. By \cite[Theorem 3.1]{krstic1998inverse}, we get that
\begin{flalign}\label{eq:example1-krstic-opt-ctrl}
    u=\alpha^*(x)=-6x^3
\end{flalign}
is an inverse optimal controller minimizing the cost functional (\ref{Ju_krstic}), with $E(x)=x^4$, $\bar{l}(x)=4x^6$, $\bar{R}_1(x)=\frac{1}{3}$, and $\gamma_0(s)=s^2$.
Moreover, the derivative of $V(x)$ along the solution of closed-loop system (\ref{counterexample}), (\ref{eq:example1-krstic-opt-ctrl}) satisfies $\dot{V}(x)\leq-4x^6+w^2$, which implies that the closed-loop system is ISS with respect to $w$. However, such a closed-loop system is not finite-gain $L_2$ stable. To see this, for the case $w\equiv0$, the solution of closed-loop system (\ref{counterexample}), (\ref{eq:example1-krstic-opt-ctrl}) is $x(t)=\frac{x_0}{\sqrt{1+10x_0^2t}}$, and thus,
\begin{align}
   \int_0^t y(\tau)^2 d\tau
    = \int_0^t \frac{x_0^2}{1+10x_0^2\tau}d\tau
    = \frac{1}{10}\ln(1+10x_0^2t)
    \nonumber
\end{align}
which implies that $y$ is not an $L_2$ signal. Therefore, the closed-loop system (\ref{counterexample}), (\ref{eq:example1-krstic-opt-ctrl}) is not finite-gain $L_2$ stable.
\end{example}

Next, we show that the resulting inverse optimal controller can guarantee both ISS and IOS, if the output is penalized.

\begin{example}\rm
Revisit the system in Example 1. Take $V(x)=\frac{1}{2}x^2$ and consider the controller
\begin{flalign}\label{eq:example2-penal-output}
    u=\alpha^*(x)=-4x^3-2.5x.
\end{flalign}
It can be verified that (\ref{eq:example2-penal-output}) is an inverse optimal controller minimizing the cost functional (\ref{Ju_de}), with $E(x)=2x^2$, $l(x)=4x^4$, $R_2(x)=1$, $R_1(x)=(2x^2+\frac{5}{4})^{-1}$ and $\gamma_0(s)=s^2$. In the following, we show that the controller (\ref{eq:example2-penal-output}) can ensure both ISS and IOS. The derivative of $V(x)$ along the solution of the closed-loop system (\ref{counterexample}), (\ref{eq:example2-penal-output}) satisfies $\dot{V}(x)\leq-3x^4-1.5x^2+w^2$, and thus,  the closed-loop system is ISS. Besides,
$y^2 + \dot{V}(x)
= y^2 -3x^4-2.5x^2+xw
\leq -3x^4-0.5x^2+w^2$,
which implies $||y||_2\leq ||w||_2+|x_0|$. Therefore, the closed-loop system (\ref{counterexample}), (\ref{eq:example2-penal-output}) is finite-gain $L_2$ stable.
\end{example}

\subsection{Main results}

Now we derive sufficient conditions for solving the inverse optimal gain assignment problem in Definition \ref{def:inv-opt}. Let $V(x)$ be a homogeneous Lyapunov candidate of degree $k+2r_0$. Because $V(x)$ may not be differentiable at $x=0$, we introduce the following lemma to ensure that the time derivative of $V(x(t))$ along the solution of system (\ref{sys}) is well-defined.

\begin{proposition}\rm
Consider system (\ref{sys}). Suppose that $V(x)$ is a homogeneous function of degree $k+2r_0$. Then the Lie derivatives $L_fV(x)$, $L_{G_1}V(x)$ and $L_{G_2}V(x)$ are well-defined on the whole domain $\R^n$.
\end{proposition}

\textbf{Proof.}
Because $V(x)$ is continuously differentiable on $\R^n\backslash\{0\}$, $L_fV(x)$ is continuous and well-defined on $\R^n\backslash\{0\}$. Moreover, since $L_fV(x)$ is homogeneous of degree $2(k+r_0)>0$, it  follows from Lemma \ref{lem:bhat-homo-cont} that $L_fV(x)$ is continuous and well-defined on whole domain $\R^n$. Similarly, because $L_{G_1}V$ and $L_{G_2}V$ are homogeneous functions of degree $k+r_0>0$, we can use Lemma \ref{lem:bhat-homo-cont} to verify that $L_{G_1}V(x)$ and $L_{G_2}V(x)$ are continuous and well-defined on $\R^n$.
\hfill $\blacksquare$

To determine the penalty $\gamma_0(|w|)$ on the disturbance, we construct an auxiliary system for (\ref{sys}) as follows:
\begin{align}\label{au_sys}
  \dot{x} = \tilde{f}(x) + G_1(x)u
\end{align}
where $\tilde{f}(x)$ is a function satisfying $\tilde{f}(0)=f(0)$ and
\begin{equation}\label{f_tilde}
    \tilde{f}(x) = f(x) + G_2(x)\ell_\gamma(2|L_{G_2}V|)\frac{L_{G_2}V^T}{|L_{G_2}V|^2},\;\;\forall x\in\R^n\backslash\{0\}
\end{equation}
Herein, $\gamma$ is a $K_\infty$-function whose derivative $\gamma^\prime$ is also of class $K_\infty$.

\begin{lemma}\label{lem:LF-homo}\rm
Suppose that $V(x)$ is homogeneous of degree $k+2r_0$, and the Legendre-Fenchel transform of $\gamma$ satisfies $\ell_\gamma(2\varepsilon s)=\varepsilon^2\ell_\gamma(2s)$ for any $ \varepsilon>0$. Then $\tilde{f}(x)$ in (\ref{f_tilde}) is continuous on $\R^n$ and homogeneous of degree $k$. Moreover, $L_{\tilde{f}}V(x)$ is well-defined on the whole domain $\R^n$.
\end{lemma}
\textbf{Proof.}
Since $V(x)$ is a homogeneous function of degree $k+2r_0$ and $G_2(x)$ is a homogenous vector field of degree $\varsigma=-r_0$, it follows that $L_{G_2}V(x)$ is a homogenous function of degree $k+r_0$ and thus $\ell_\gamma(2 |L_{G_2}V(\Delta_\varepsilon^r x)|)=\ell_\gamma(2\varepsilon
^{k+r_0}|L_{G_2}V(x)|)$. Given that $\ell_\gamma(2\varepsilon s)=\varepsilon^2\ell_\gamma(2s)$, we have $\ell_\gamma(2\varepsilon
^{k+r_0}|L_{G_2}V(x)|)=\varepsilon
^{2(k+r_0)}\ell_\gamma(2|L_{G_2}V(x)|)$, namely, $\ell_\gamma(2|L_{G_2}V|)$ is homogeneous of degree $2(k+r_0)$. Because the degree of $|L_{G_2}V(x)|^2$ is $2(k+r_0)$, $G_2(x)\ell_\gamma(2|L_{G_2}V|)\frac{L_{G_2}V^T}{|L_{G_2}V|^2}$ is a homogenous vector field of degree $k$. Therefore, $\tilde{f}(x)$ in (\ref{f_tilde}) is a homogeneous vector field of degree $k$. Because $V(x)$ is homogeneous and positive definite, $L_{G_2}V(x)$ is continuous on $\R^n\backslash\{0\}$. Also note that $\ell_\gamma(2|L_{G_2}V|)\frac{L_{G_2}V^T}{|L_{G_2}V|^2}$ is homogeneous of degree $k+r_0>0$. Hence, it follows from Lemma \ref{lem:bhat-homo-cont} that $\ell_\gamma(2|L_{G_2}V|)\frac{L_{G_2}V^T}{|L_{G_2}V|^2}$ is continuous on the whole domain $\R^n$, implying that $\tilde{f}(x)$ is also continuous globally. Similarly, we can verify that $L_{\tilde{f}}V(x)$ is well-defined on the whole domain $\R^n$.
\hfill $\blacksquare$

\begin{lemma}\label{lem:func-H}\rm
Consider the functions $V$ and $\gamma$ as in Lemma \ref{lem:LF-homo}. Suppose that there exists a well-defined homogeneous function $R(x)>0$ of degree zero with respect to $\Delta_\varepsilon^r$, where $|R(x)|$ is bounded away from zero, such that
\begin{align}\label{au_u}
  u = \alpha(x)= -\frac{\kappa}{2\vartheta^2}R(x)^{-1}(L_{G_1}V)^T,\quad \kappa>\kappa_0>0
\end{align}
asymptotically stabilizes auxiliary system (\ref{au_sys}) at $x=0$. Then, for the function $H_\kappa(x)$ satisfying $H_k(0)=0$ and
\begin{flalign}\label{lkappa}
  H_\kappa(x)
  &=
  -\kappa\Big[L_{\tilde{f}}V(x)
                    +L_{G_1}V(x)\alpha(x)\Big]
    \nonumber\\
  &\;\;\;\;\;\;\;\;\;\;\;\;
    - h(x)^TR(x)h(x),\;\;\;\;\;\;\forall x\in\R^n\backslash\{0\},
\end{flalign}
there is a constant $\kappa$ such that $H_\kappa(x)$ is positive definite on $\R^n$.
\end{lemma}
\textbf{Proof.}
Since the control law (\ref{au_u}) stabilizes (\ref{au_sys}), by \cite[Lemmas 4.1-4.2]{Bhat_mcss_2005}, there exists a constant $c_1>0$ such that
\begin{equation}\label{proper_dvf}
    L_{\tilde{f}}V(x)+L_{G_1}V(x)\alpha(x)
  \leq -c_1\Gamma(x)^{2(k+r_0)},\;\;\forall x\in\R^n.
\end{equation}
Recalling $h(0)=0$, it follows from (\ref{lkappa})-(\ref{proper_dvf}) that $\lim_{x\rightarrow0}H_k(x)=0$. Hence, $H_k(x)$ is continuous on $\R^n$. The rest is to show that $H_\kappa(x)>0$ holds on $\R^n\backslash\{0\}$. Introduce the unity homogeneous sphere
\begin{align}\label{sphere}
    S = \{x: \Gamma(x)=1\}.
\end{align}
In order to utilize homogeneity properties to show $H_\kappa(x)>0$ for all $x\in\R^n\backslash\{0\}$, we have the following two steps.

\emph{Step 1: Prove that $H_\kappa(x)$ is a homogeneous function.} Because $G_1(x)$ is a homogeneous vector field of degree $\varsigma=-r_0$, and $R(x)$ is a homogeneous function of degree zero, the control law $u=\alpha(x)$ in (\ref{au_u}) is homogenous of degree $k+r_0$. By Lemma \ref{lem:LF-homo}, $\tilde{f}(x)$ is homogeneous of degree $k$, and therefore, $L_{\tilde{f}}V(x)+L_{G_1}V(x)\alpha(x)$ is homogeneous of degree $2(k+r_0)$. Since $h(x)$ is a homogeneous function of degree $k+r_0$, we obtain that $h^TR(x)h$ is a homogeneous function of degree $2(k+r_0)$. Therefore, $H_\kappa(x)$ is a homogenous function of degree $2(k+r_0)$.

\emph{Step 2: Prove $H_\kappa(x)>0$ for all $x\in S$}. Define the sets
\begin{align}
  P^{+} & = \Big\{x\in \R^n \backslash \{0\}: \;\;\; L_{\tilde{f}}V(x)\geq0\Big\},  \\
  P^{-} & = \Big\{x\in \R^n: \;\;\; L_{\tilde{f}}V(x)<0\Big\},  \\
  P^{0} & = \Big\{x\in \R^n \backslash \{0\}: \;\;\;  L_{G_1}V(x)=0\Big\}.
\end{align}
From (\ref{proper_dvf}), it gives that $P^{0}\subset P^{-}$. Thus, there exists an open set $Q_0$ such that $P^0\cup\{0\}\subset Q_0\cup\{0\}$ and $\cl(Q_0)\subset P^-\cup\{0\}$. For any $x\in S$, we prove $H_\kappa(x)>0$ in two cases:

(i) $x\in S\cap \cl{(Q_0)}$. Since $S$ and $\cl{(Q_0)}$ are closed sets, there are constants $\rho_1$ and $\rho_2$ such that
\begin{align}\label{}
  0 & < \rho_1 \leq -\max_{x\in S\cap \cl{(Q_0)}} L_{\tilde{f}}V(x), \\
  0 & \leq \max_{x\in S\cap \cl{(Q_0)}}h(x)^TR(x)h(x) \leq \rho_2.
\end{align}
Then taking $\kappa_0\geq\kappa_c$, where $\kappa_c=\rho_2/\rho_1$, we have
\begin{equation}\label{case1_kappah}
    \kappa L_{\tilde{f}}V(x)
  + h(x)^TR(x)h(x)<0,\;\;\forall x\in S\cap \cl(Q_0).
\end{equation}
With (\ref{au_u}), (\ref{lkappa}) and (\ref{case1_kappah}), we have $H_\kappa(x)>0$ on $S\cap \cl(Q_0)$.

(ii) $x\in S\backslash Q_0$. Since $S\backslash Q_0$ is a closed set, the term $L_{G_1}VR(x)^{-1}(L_{G_1}V)^T$ has a nonzero minimum and $h(x)^TR(x)h(x)$ has a maximum on $S\backslash Q_0$. Then there exist real numbers $\rho_3>0$ and $\rho_4>0$ such that
\begin{align}
  0 & <\rho_3\leq \min_{x\in S\backslash Q_0}L_{G_1}VR(x)^{-1}(L_{G_1}V)^T, \label{case2_leq1} \\
  0 & \leq \max_{x\in S\backslash Q_0} h(x)^TR(x)h(x) \leq \rho_4. \label{case2_leq2}
\end{align}
Denote the maximum of $L_{\tilde{f}}V(x)$ by $\rho$. With combination of (\ref{au_u})-(\ref{lkappa}) and (\ref{case2_leq1})-(\ref{case2_leq2}), it yields  that
\begin{align}\label{}
  -H_\kappa(x)
  \leq \kappa\rho + \rho_4 - \frac{\kappa^2}{2\vartheta^2}\rho_3,\;\;
  \forall x\in S\backslash Q_0.
\end{align}
Notice that the solutions $\kappa_1$ and $\kappa_2$ of the quadratic equation $\kappa\rho + \rho_4 - \frac{\kappa^2}{2\vartheta^2}\rho_3=0$ are computed as $\kappa_1=\vartheta^2(\rho+\sqrt{\rho^2+2\rho_3\rho_4/\vartheta^2})/\rho_3$ and $\kappa_2=\vartheta^2(\rho-\sqrt{\rho^2+2\rho_3\rho_4/\vartheta^2})/\rho_3$. Taking  $\kappa_0\geq \kappa_1$, it yields that $H_\kappa(x)>0$ on {$S\backslash Q_0$}.

By taking $\kappa>\kappa_0\geq\max\{\kappa_c, \ \kappa_1\}$, one can ensure $H_\kappa(x)>0$ for all $x\in S$. Then it follows from homogeneity properties that $H_\kappa(x)>0$ holds on $\R^n\backslash\{0\}$. In summary, $H_\kappa(x)$ is positive definite on $\R^n$.
\hfill $\blacksquare$

With Lemma \ref{lem:LF-homo} and Lemma \ref{lem:func-H}, we have the following result.

\begin{theorem}\label{thm:homo-gain-assign}\rm
Consider system (\ref{sys}) with the functions $V$, $\gamma$, $R$ and constant $\kappa_0$ given by Lemma \ref{lem:LF-homo} and Lemma \ref{lem:func-H}. If the control law $u=\alpha(x)$ in (\ref{au_u}) asymptotically stabilizes auxiliary system (\ref{au_sys}) at $x=0$, then under the control law
\begin{align}\label{optimal_u}
  u = \alpha^*(x)= -\frac{\beta\kappa}{2\vartheta^2}R(x)^{-1}(L_{G_1}V)^T, \quad \beta\geq2
\end{align}
the following properties hold:
\begin{basedescript}{\desclabelstyle{\pushlabel}\desclabelwidth{0.7cm}}
\item[\hspace{0.17cm}(i)] The cost functional (\ref{Ju_de}) is minimized by (\ref{optimal_u}) with $E(x(t))=2\beta V(x(t))$, $R_1(x)=\frac{\vartheta^2}{\kappa}R(x)$, $R_2(x)=\frac{1}{\kappa}R(x)$, $\gamma_0(|w|)=\beta\lambda\gamma(\frac{|w|}{\lambda})$,
and
\begin{flalign}
    l(x)=\bar{l}(x)-\frac{h(x)^TR(x)h(x)}{\kappa},
        \label{lx}
\end{flalign}
where $\bar{l}(x)$ is a function satisfying $\bar{l}(0)=0$ and
\begin{flalign}
    &\bar{l}(x)=-2\beta\Big(L_{\tilde{f}}V + L_{G_1}V\alpha(x)\Big)
    +\beta(2-\lambda)\ell_\gamma(2|L_{G_2}V|)
        \nonumber\\
    &\;\;\;\;
    -\beta(\beta-2)L_{G_1}V\alpha(x),\;\;\;\;\forall x\in\R^n\backslash\{0\}.
        \label{eq:bar-lx}
\end{flalign}
Herein, $0<\lambda\leq2$ and $\beta\geq2$ are constants.
\item[\hspace{0.17cm}(ii)] The closed-loop system (\ref{sys}), (\ref{optimal_u}) is ISS.
\item[\hspace{0.17cm}(iii)] The closed-loop system (\ref{sys}), (\ref{optimal_u}) is finite-gain $L_2$ stable, if one picks $\gamma(s)=\frac{1}{\mu}s^2$, $\mu>0$.
\end{basedescript}
\end{theorem}
\textbf{Proof.} \textbf{(i)} Because $\bar{l}(x)$ is homogeneous of degree $2(k+r_0)>0$, we have $\lim_{\varepsilon\rightarrow0}\bar{l}(\varepsilon^{r_1}x_1,\ldots,\varepsilon^{r_n}x_n)
    =\lim_{\varepsilon\rightarrow0}\varepsilon^{2(k+r_0)}\bar{l}(x)
    =0$,
and thus, $\bar{l}(x)$ is continuous on $\R^n$. Due to $h(0)=0$, it yields that $l(0)=0$. Furthermore, with $\beta\geq2$, $0<\lambda\leq2$, and $\kappa>0$, it follows from (\ref{optimal_u})-(\ref{eq:bar-lx}) that
\begin{flalign}
    l(x)
    \geq2\beta\Bigg[-\Big(L_{\tilde{f}}V + L_{G_1}V\alpha(x)\Big)
    -\frac{h(x)^TR(x)h(x)}{2\beta\kappa}\Bigg]
    \geq \frac{2\beta H_\kappa(x)}{\kappa}
    \nonumber
\end{flalign}
which, together with Lemma \ref{lem:func-H}, implies that $l(x)$ is positive definite.
Since $l(x)$ is also a homogeneous function, we obtain from \cite[Lemma 4.1]{Bhat_mcss_2005} that $l(x)$ is radially unbounded. Therefore, the cost functional $J(u)$ is meaningful. By (\ref{control_gain}), we have $y^TR_2(x)y = h(x)^TR_2(x)h(x) + \vartheta^2u^TR_2(x)u$.
By combining this with (\ref{lx}) and (\ref{eq:bar-lx}), it gives that
\begin{flalign}\label{eq:new-lx-uSq-ySq}
    l(x)+u^TR_1u+y^TR_2y=\bar{l}(x)+\frac{2\vartheta^2}{\kappa}u^TRu.
\end{flalign}
Substituting (\ref{lx}), (\ref{eq:bar-lx}) and (\ref{eq:new-lx-uSq-ySq}) into (\ref{Ju_de}), we get
\begin{align}\label{Ju1}
  & J(u) \nonumber \\
  & = \sup_{w\in\Omega_w}\bigg\{2\beta V(x(\infty)) -2\beta \int_{0}^{\infty}\bigg(L_fV + L_{G_1}Vu
  + L_{G_2}Vw\bigg)dt
  \nonumber \\
  & \quad
  + \int_{0}^{\infty}\bigg(\frac{2\vartheta^2}{\kappa}u^TRu
  + 2\beta L_{G_1}Vu
  + \frac{\kappa\beta^2}{2\vartheta^2}L_{G_1}VR^{-1}(L_{G_1}V)^T \bigg)dt\nonumber \\
  & \quad
  + \int_{0}^{\infty}\underbrace{\beta\bigg( 2 L_{G_2}Vw
  - \lambda\ell_\gamma(2|L_{G_2}V|)
  - \lambda\gamma\Big(\frac{|w|}{\lambda}\Big)\bigg)}_{\Delta_w}dt
  \bigg\}
  \nonumber \\
  & = 2\beta V(x(0))
  + \frac{2\vartheta^2}{\kappa}\int_{0}^{\infty}(u-\alpha^*)^TR(u-\alpha^*)dt
  \nonumber \\
  & \quad
  + \sup_{w\in\Omega_w}\bigg\{\int_{0}^{\infty}\Delta_wdt\bigg\}.
\end{align}
Now we analyze the term $\Delta_w$. By (\ref{young_gamma}), one has $2L_{G_2}V w\leq \lambda\ell_\gamma(2|L_{G_2}V|)
  + \lambda\gamma\Big(\frac{|w|}{\lambda}\Big)$,
which yields $\Delta_w\leq0$ and thus $\sup_{w\in\Omega_w}\int_{0}^{\infty}\Delta_wdt=0$. Thus, $J(u)$ achieves its minimum at $u=\alpha^*(x)$ and the minimum is $J(u)_{\min}=2\beta V(x(0))$.

\textbf{(ii)} With (\ref{proper_dvf}), the derivative of $V(x)$ along the solution of the closed-loop system (\ref{sys}), (\ref{optimal_u}) is
\begin{align}\label{dv_sysiss}
  \dot{V}(x) & =
  L_fV+L_{G_1}V\alpha^*(x)+L_{G_2}Vw
  \nonumber\\
  & \leq -c_1\Gamma(x)^{2(k+r_0)} -\ell_\gamma(2|L_{G_2}V|) +L_{G_2}Vw
  \nonumber \\
  & \quad +(1-\beta)\frac{\kappa}{2\vartheta^2}L_{G_1}VR^{-1}(L_{G_1}V)^T
  \nonumber \\
  & \leq -c_1\Gamma(x)^{2(k+r_0)} + \gamma\Big(\frac{|w|}{2}\Big).
\end{align}
Recall that $\Gamma(x)$ is positive definite radially unbounded, and thus, the closed-loop system (\ref{sys}), (\ref{optimal_u}) is ISS.

\textbf{(iii)} With the derivative of $V(x)$ along the solution of the closed-loop system (\ref{sys}), (\ref{optimal_u}), and by (\ref{control_gain}), we have
\begin{align}\label{dv_sysios_case2}
  \kappa \dot{V}(x) + \frac{y^TRy}{\beta}
  & = \kappa
  \Big(L_fV(x)+L_{G_1}V(x)\alpha^*(x)+L_{G_2}V(x)w\Big)
  \nonumber \\
  & \quad
  + \frac{h^TRh}{\beta}
  + \frac{\vartheta^2\alpha^{*T}R\alpha^*}{\beta}
  \nonumber \\
  & \leq -H_\kappa(x) + \kappa L_{G_2}Vw
  - \kappa\ell_\gamma(2|L_{G_2}V|)
  \nonumber \\
  & \quad + \Big(1-\frac{\beta}{2}\Big)\frac{\kappa^2}{2\vartheta^2}L_{G_1}VR^{-1}(L_{G_1}V)^T
  \nonumber \\
  & \leq -H_\kappa(x)
  + \kappa\gamma\Big(\frac{|w|}{2}\Big).
\end{align}
Denote $\rho_m=\inf_{x\in\R^n}|R(x)|$. Since $|R(x)|$ is bounded away from zero, $\rho_m^{-1}$ is well-defined. By Lemma \ref{lem:func-H}, we have $H_\kappa(x)\geq0$. Then from (\ref{dv_sysios_case2}), it further yields $y^Ty \leq \rho_m^{-1}\kappa\beta\gamma\Big(\frac{|w|}{2}\Big)
  - \rho_m^{-1}\kappa\beta\dot{V}(x)$.
Since $\gamma(s)=\frac{1}{\mu}s^2$, we have $\int_{0}^{T} y^Tydt
  \leq \int_{0}^{T} \frac{\rho_m^{-1}\kappa\beta}{4\mu}w^Twdt
  + \rho_m^{-1}\kappa\beta V(x(0))$
for all $T>0$. Therefore, there exist a finite gain $\kappa_L=\sqrt{\kappa\beta/(4\rho_m\mu)}$ and a constant $c_0=\sqrt{\rho_m^{-1}\kappa\beta V(x(0))}$, which depends on only the initialization $x(0)$, such that $||y||_2\leq \kappa_L||w||_2
  + c_0$.
The proof is completed. \hfill $\blacksquare$

\begin{remark}\rm
From Theorem 1, we see that putting an additional penalty on the output offers the advantage of providing provable guarantees for both ISS and IOS. However, it also complicates the construction of a meaningful cost functional $J(u)$. Moreover, the methods in \cite{krstic1998inverse} cannot be directly used to solve our inverse optimal control problem, because their cost functional cannot be converted to ours. To see this, rewrite (\ref{Ju_krstic}) as
\begin{align}\label{Ju_krstic-rewrite}
  J(u) & = \sup_{w\in \Omega_w}\bigg\{ \lim_{t\rightarrow\infty}\bigg[
  E(x(t))
  + \int_{0}^{t}\bigg(\tilde{l}(x)
  \nonumber \\
  & \quad \quad
   + u^T\tilde{R}_1(x)u
   +y^TR_2(x)y
   -\gamma_0(|w|)
   \bigg)d\tau
   \bigg]\bigg\},
\end{align}
where $\tilde{l}(x)=\bar{l}(x)-h(x)^TR_2(x)h(x)$ and $\tilde{R}_1(x)=\bar{R}_1(x)-\vartheta^2R_2(x)$. Obviously, $\tilde{l}(x)$ may not be positive definite, which violates the requirement of state penalty. In contrast to \cite{krstic1998inverse}, the main challenge lies in the construction of $l(x)$ and $\gamma(s)$:
\begin{itemize}
  \item The difficulty of constructing $l(x)$ lies in that it is hard to ensure $l(x)>0$ on $\R^n\backslash\{0\}$. Specifically, as seen in (\ref{lx}), both $\bar{l}(x)$ and $h(x)^TR(x)h(x)$ are positive and nonlinear, it is hard to choose a $\kappa$ to make the inequality $\bar{l}(x)>\frac{h(x)^TR(x)h(x)}{\kappa}$ hold for all $x\in\R^n\backslash\{0\}$. To handle this issue, we provide a technique in Lemma \ref{lem:func-H} for the selection of $\kappa$, based on the ability of homogeneity to extend local properties to global ones.
  \item The difficulty of constructing $\gamma(s)$ lies in that its selection directly affects proving the positive definiteness of $l(x)$. Specifically, Lemma \ref{lem:func-H} requires that the auxiliary system (\ref{au_sys}) is homogeneous. However, as seen in (\ref{au_sys}) and (\ref{f_tilde}), the auxiliary system may lose homogeneity, due to the term $\ell_\gamma(2|L_{G_2}V|)$. To this end, we provide a method in Lemma \ref{lem:LF-homo} for choosing a $\gamma(s)$ to ensure that the auxiliary system is homogeneous.
\end{itemize}
\end{remark}

\begin{remark}\rm
The auxiliary system (\ref{au_sys}) is used to construct an inverse optimal controller minimizing the cost functional with a penalty on the disturbance $w$. This idea was originally proposed in \cite{krstic1998inverse}. Although \cite{Andrieu_V_2008,Bernuau_2013,Bernuau_2014, Nakamura2009tac} have studied ISS analysis of homogeneous system $\dot{x}=f(x,w)$ or inverse optimal stabilization of disturbance-free homogeneous control system $\dot{x}=f(x)+G_1(x)u$, the combination of these results cannot be used to construct a homogeneous inverse ISS controller minimizing the cost functional of \cite{krstic1998inverse}. This is because it remains unclear how to utilize these results to construct a homogeneous auxiliary system. Furthermore, as discussed in Remark 2, proving the positive definiteness of our state penalty $l(x)$ relies on homogeneity of the auxiliary system. Therefore, a simple combination of the analysis in \cite{Andrieu_V_2008,Bernuau_2013,Bernuau_2014, Nakamura2009tac, krstic1998inverse} cannot establish Theorem 1. As shown in Example 3 below, even though one can construct a homogeneous inverse optimal controller minimizing the cost functional of \cite{krstic1998inverse}, it may be impossible for this controller to minimize our cost functional.
\end{remark}

\begin{example}\rm
Consider the system
\begin{align}\label{counterexample2}
    \dot{x}_1  = -x_1^3 + x_2^3,\;\;
    \dot{x}_2  = u+w,\;\;
    y & = x_2^3
\end{align}
which is homogeneous of degree $2$ with respect to the dilation $(1,1)$. Herein, $f(x)=[-x_1^3 + x_2^3,0]^T$, $G_1(x)=G_2(x)=[0,1]^T$, $h(x)=x_2^3$, $d=0$ and system (\ref{counterexample2}) is the same form of (\ref{sys}) with $k=2$, $\varsigma=-1$, $r_0=1$. Take $V(x)=\frac{1}{4}x_1^4+\frac{1}{4}x_2^4$, which is homogeneous of degree $4$. Pick a $K_\infty$-function $\rho(s)=s^{1/3}$. Then design the controller as
\begin{align}\label{alphac_ex2}
    u=\alpha(x)
    = \left\{
  \begin{array}{rll}
    & -\frac{\varpi+\sqrt{\varpi^2+x_2^{12}}}{x_2^3}, & L_{G_1}V \neq 0 \\
    & 0, & L_{G_1}V = 0
    \end{array} \right.
\end{align}
where $\varpi=L_fV+|L_{G_2}V|\rho^{-1}(|x|)=-x_1^6+x_1^3x_2^3+|x_2|^3
(x_1^2+x_2^2)^{3/2}$. It is easy to verify that $\alpha(x)$ is homogeneous of degree 3. The derivative of $V(x)$ along the solution of the closed-loop system (\ref{counterexample2}) and $u=\frac{1}{2}\alpha(x)$ satisfies $\dot{V}\leq-\frac{1}{2}\big[-\varpi+(\varpi^2+x_2^{12})^{1/2}\big]
-|x_2|^3\big[\rho^{-1}(|x|)-|w|\big]$. Thus, $u=\frac{1}{2}\alpha(x)$ is an ISS controller. Let $\gamma(s)=s^2$ and by \cite[Theorem 3.1]{krstic1998inverse}, $u=\alpha(x)$ is an inverse optimal controller minimizing the cost functional (\ref{Ju_krstic}) with $E(x)=x_1^4+x_2^4$, $\bar{l}(x) = 2\big[-\varpi+(\varpi^2+x_2^{12})^{1/2}\big]
    + 4\big[|x_2|^3(x_1^2+x_2^2)^{3/2}-x_2^6 \big]$, $\bar{R}_1(x)=2x_2^6/[\varpi+(\varpi^2+x_2^{12})^{1/2}]$, and $\gamma_0(s)=s^2$. Clearly, $\bar{l}(x)$ is positive definite. However, $u=\alpha(x)$ is not inverse optimal with respect to the cost functional (\ref{Ju_de}) because based on the same design, the penalty on the state, denoted by $\tilde{l}(x)$, is $\tilde{l}(x) = \bar{l}(x)-y^TR_2(x)y$, which may be negative for some $x\in\R^n$. For example, suppose that $R_2(x)=1$, then
$\tilde{l}(x)
    = 2\big(-\varpi+(\varpi^2+x_2^{12})^{1/2}\big)
    + 4\big(|x_2|^3(x_1^2+x_2^2)^{3/2}-x_2^6  \big)
    -x_2^6$
is not positive definite, violating the requirement of the state penalty.
\end{example}

\begin{remark}\rm
The gain margin property is a benefit of inverse optimal controllers. In \cite{Nakamura2009tac}, it is shown that a tunable gain margin $(\frac{k+\sigma-s}{k+\sigma},\infty)$ can be achieved by employing a polynomial control penalty $\frac{s}{k+\sigma}R_1^{\frac{k+\sigma-s}{s}}(x)
|u|^{\frac{k+\sigma}{s}}$, where $\sigma$ and $s$ are the homogeneous degrees of the control Lyapunov function and the inverse optimal controller, respectively. Our control penalty is quadratic and can be viewed as a special case of \cite{Nakamura2009tac}, by taking $\sigma=k+2r_0$ and $s=k+r_0$. However, our inverse optimal controller also possesses tunable gain margin $(1/\beta,\infty)$. This can be verified by the following observations.
\begin{itemize}
  \item Increasing the gain of our inverse optimal controller (\ref{optimal_u}) implies that its ``stabilizing effort'' is amplified. As a result, it is reasonable that such a controller can tolerate infinite gain increase.
  \item By comparing (\ref{au_u}) and (\ref{optimal_u}), we see that our inverse optimal controller (\ref{optimal_u}) has $\beta$ times more control effort than is required to stabilize the control system. Thus, it is natural that this controller can tolerate $1/\beta$ gain reduction.
\end{itemize}
\end{remark}

In Theorem 1, we derive sufficient conditions on solving the inverse optimal gain assignment problem of homogeneous system (\ref{sys}). In the following, we show how to design inverse optimal controllers that satisfy the conditions in Theorem 1.

\begin{theorem}\rm
If system (\ref{sys}) is homogeneously stabilizable in the case of $w\equiv0$, i.e., the system
\begin{align}\label{sys_wzero_y}
  \dot{x} & = f(x) + G_1(x)u
\end{align}
is homogeneously stabilizable, then the inverse optimal gain assignment problem for (\ref{sys}) is solvable.
\end{theorem}
\textbf{Proof.}
Suppose that system (\ref{sys_wzero_y}) is homogeneously stabilizable by $u=\alpha_h(x)$. Then, by \cite[Theorem 2 and Proposition 2]{Rosier1992scl}, there exists a homogeneous Lyapunov function $V(x)\in C^\infty(\R^n\backslash\{0\})$ of degree $k+2r_0$ such that
\begin{align}\label{dv_sys_zerow}
  L_fV + L_{G_1}V\alpha_h(x) < 0, \quad \forall x\in\R^n\backslash\{0\}
\end{align}
With homogeneity properties of $V(x)$, it follows from \cite[Lemma 4.2]{Bhat_mcss_2005} that
\begin{align}
  L_fV + L_{G_1}V\alpha_h(x) & \leq -c_1\Gamma(x)^{2(k+r_0)}, \label{dv_sys_zerow_Ix1}
   \\
  |L_{G_2}V|& \leq c_2\Gamma(x)^{k+r_0}, \label{dv_sys_zerow_Ix2}
\end{align}
where $c_1, c_2$ are positive constants such that $c_1=\min\limits_{\{s:|\Gamma(s)|=1\}} [-L_fV - L_{G_1}V\alpha_h(s)]$ and $c_2=\max\limits_{\{s:|\Gamma(s)|=1\}} |L_{G_2}V(s)|$. 
Introduce the Sontag-type controller
\begin{align}\label{alphas}
  \alpha_s(x)= \left\{
    \begin{array}{rll}
    & -\bigg[
    c_{10}
    +
    \frac{\phi + \sqrt{\phi^2 + [L_{G_1}V(L_{G_1}V)^T]^2}}{L_{G_1}V(L_{G_1}V)^T}\bigg](L_{G_1}V)^T, & (L_{G_1}V)^T \neq 0 \\
    & 0, & (L_{G_1}V)^T = 0
    \end{array} \right.
\end{align}
where $c_{10}>0$ is a real number, and $\phi(x)$ is a function satisfying $\phi(0)=0$ and
\begin{align}\label{phi}
  \phi(x) = L_fV + |L_{G_2}V|\pi\Big(\Gamma(x)\Big),\;\;\;\;\forall x\in\R^n\backslash\{0\}.
\end{align}
Herein, $\pi(s)=c_6s^{k+r_0}$, where $c_6>0$ is a constant defined after (\ref{leadto_ho3}). Because $L_fV(x)$ is homogeneous of degree $2(k+r_0)$, and $L_{G_2}V$ and $\pi(\Gamma(x))$ are homogenous of degree $k+r_0$, $\phi(x)$ is homogeneous of degree $2(k+r_0)>0$. With homogeneity, we have
$\lim_{\varepsilon\rightarrow0}\phi(\varepsilon^{r_1}x_1,\ldots,\varepsilon^{r_n}x_n)
    =\lim_{\varepsilon\rightarrow0}\varepsilon^{2(k+r_0)}\phi(x)
    =0$. Hence, $\phi(x)$ is continuous on $\R^n$.
Clearly,
\begin{align}\label{eq:thm-two-alpha}
  u=\alpha(x)= \frac{\kappa}{2}\alpha_s(x)
\end{align}
is of the form (\ref{au_u}) with
\begin{align}\label{Rx}
  R(x) = \left\{
  \begin{array}{rll}
    & \frac{1}{\vartheta^2}\Bigg[
    c_{10}+
    \frac{\Big(\phi + \sqrt{\phi^2 + [L_{G_1}V(L_{G_1}V)^T]^2}\Big)}{ L_{G_1}V(L_{G_1}V)^T}\Bigg]^{-1}, & (L_{G_1}V)^T \neq 0 \\
    & \frac{1}{\vartheta^2c_{10}}, & (L_{G_1}V)^T = 0
    \end{array} \right.
\end{align}
Obviously, $R(x)>0$. With the homogeneity degrees of $\phi(x)$ and $L_{G_1}V(x)$, we can compute that $R(x)$ is homogeneous of degree zero and $u=\alpha(x)= \frac{\kappa}{2}\alpha_s(x)$ is homogeneous of degree $k+r_0$.

Now, we are ready to show that the weight matrix $R(x)$ and the controller $u=\alpha(x)=\frac{\kappa}{2}\alpha_s(x)$ are continuous on $\R^n\backslash\{0\}$. As proved in \cite{Sontag_scl_1989}, $\alpha_s(x)$ is smooth on $\R^n\backslash\{0\}$, provided that
\begin{align}\label{LG2Vleadtophizero}
  L_{G_1}V=0 \;\; \Longrightarrow
  \;\;
  \phi<0.
\end{align}
In the following, we show that (\ref{LG2Vleadtophizero}) is always true, if (\ref{sys_wzero_y}) is homogeneously stabilizable. By (\ref{dv_sys_zerow_Ix1}) and (\ref{dv_sys_zerow_Ix2}), the derivative of $V(x)$ along the solution of the closed-loop system (\ref{sys}) and $u=\alpha_h(x)$ is
\begin{align}\label{dv_sys}
  L_fV + L_{G_1}V\alpha_h(x) + L_{G_2}Vw
  & \leq -c_1\Gamma(x)^{2(k+r_0)} + c_2\Gamma(x)^{k+r_0}|w|
  \nonumber \\
  & \leq -c_3\Gamma(x)^{2(k+r_0)} + c_4|w|^2,
\end{align}
where $c_3=c_1/2$ and $c_4=2c_2^2/c_1$. Form (\ref{dv_sys}), we have
\begin{align}\label{leadto_ho1}
  \Gamma(x)^{k+r_0}  \geq c_5|w| \
\Rightarrow \
  L_fV + L_{G_1}V\alpha_h(x) + L_{G_2}Vw<0
\end{align}
for $x\neq0$, where $c_5=\sqrt{c_4/c_3}$. From (\ref{leadto_ho1}), if $L_{G_1}V=0$,
\begin{align}\label{leadto_ho2}
  \Gamma(x)^{k+r_0}  \geq c_5|w| \
\Rightarrow \
  L_fV + L_{G_2}Vw<0.
\end{align}
Consider a particular input  $w=\frac{(L_{G_2}V)^T}{|L_{G_2}V|}\frac{\Gamma(x)^{k+r_0}}{c_5}$, which satisfies the left part of the implication (\ref{leadto_ho2}), and substituting it into the right part of (\ref{leadto_ho2}), it yields that
\begin{align}\label{leadto_ho3}
  L_fV + L_{G_2}Vw & = L_fV + c_6|L_{G_2}V|\Gamma(x)^{k+r_0}<0,
\end{align}
where $c_6=1/c_5$. Combing with (\ref{phi}) and (\ref{leadto_ho3}), and recalling $\pi(s)=c_6s^{k+r_0}$, we conclude that, if $x\neq0$ and $L_{G_1}V=0$, then
\begin{align}\label{impli_hold}
  \phi = L_fV + |L_{G_2}V|\pi\Big(\Gamma(x)\Big)<0.
\end{align}
Thus, the implication (\ref{LG2Vleadtophizero}) holds and the control law $u=\alpha(x)$ is continuous on $\R^n\backslash\{0\}$. Because $\phi$ and $L_{G_1}V$ are continuous, $R(x)$ is continuous on $\{x\in \R^n|L_{G_1}V(x)\neq0\}$. By (\ref{leadto_ho1}) and the design of $\phi$ in (\ref{phi}), we have $\phi + L_{G_1}V\alpha_h(x)<0$ for all $x\in\R^n\backslash\{0\}$. Hence, $\phi<0$ in a sufficient small neighborhood of $\{x\in \R^n\backslash\{0\}|L_{G_1}V=0\}$, and within this region, $R(x)=1/(\vartheta^2c_{10})$. Therefore, $R(x)$ is continuous on $\R^n\backslash\{0\}$.


Now we prove that $|R(x)|$ is bounded away from zero. Recalling (\ref{leadto_ho1}), consider the particular input $w=\frac{(L_{G_2}V)^T}{|L_{G_2}V|}\frac{\Gamma(x)^{k+r_0}}{c_5}$, which satisfies the left part of the implication (\ref{leadto_ho1}), and substituting it into the right part of (\ref{leadto_ho1}), and with (\ref{phi}), we have
\begin{align}\label{R1_bound}
    L_fV + c_6|L_{G_2}V|\Gamma(x)^{k+r_0} + L_{G_1}V\alpha_h(x)<0, \quad \forall x\neq0
\end{align}
which implies that $|\phi| \leq |L_{G_1}V||\alpha_h(x)|$ for $\phi\geq0$. From (\ref{LG2Vleadtophizero}), we see that if $\phi\geq0$, then $L_{G_1}V\neq0$. Since $\alpha_h(x)$ and $L_{G_1}V$ are continuous homogeneous functions of degrees $k+r_0>0$, by \cite[Lemma 4.2]{Bhat_mcss_2005}, $|\alpha_h|\leq c_9|L_{G_1}V|$ holds with $c_9=\max\limits_{\{s:\Gamma(s)=1\}} |\alpha_h(s)|/\min\limits_{\{s:\Gamma(s)=1\}} |L_{G_1}V(s)|$ for $\phi\geq0$. Following \cite{Sontag_scl_1989}, along with (\ref{Rx}) and (\ref{R1_bound}), we have $|R^{-1}(x)|\leq\vartheta^2(c_{10}+\frac{2|\phi|}{|L_{G_1}V|^2}+1)
\leq\vartheta^2(c_{10}+2c_9+1)$ when $\phi\geq0$ and $|R^{-1}(x)|\leq\vartheta^2(c_{10}+1)$ when $\phi<0$. Thus, $|R(x)^{-1}|$ is bounded on $\R^n$, namely, $|R(x)|$ is bounded away from zero.

Afterwards, we show how to design a $K_\infty$-function $\gamma$ such that $\ell_\gamma(2\varepsilon s)=\varepsilon^2\ell_\gamma(2s)$ for any $\varepsilon>0$. Since $|L_{G_2}V|$ is homogeneous with degree of $k+r_0>0$, then by \cite[Lemma 4.2]{Bhat_mcss_2005}, we have
\begin{align}\label{c1Ixc2}
  |L_{G_2}V|\leq c_8\Gamma(x)^{(k+r_0)}
\end{align}
where $c_8=\max\limits_{\{s:\Gamma(s)=1\}} |L_{G_2}V(s)|$. Let $c_7=(1/c_8)^{1/(k+r_0)}$.
Take $\eta(2s) = s\pi\Big(c_7 s^{\frac{1}{k+r_0}}\Big)$.
Obviously, $\eta$ and its derivative $\eta^\prime$ are of class $K_\infty$. Choose $\gamma=\ell_\eta$. By Lemma \ref{lem:LF-transform}, we have $\ell_\gamma=\ell\ell_\eta=\eta$, and thus,
\begin{align}\label{lgamma}
  \ell_\gamma(2s) = s\pi\Big(c_7 s^{\frac{1}{k+r_0}}\Big)
\end{align}
which implies
\begin{align}\label{lgammasigmas}
  \ell_\gamma(2\varepsilon s)
  & = (\varepsilon s) \pi\Big(c_7 (\varepsilon s)^{\frac{1}{k+r_0}}\Big)
  \nonumber \\
  & = (\varepsilon s)\cdot c_6c_7^{k+r_0}\varepsilon s
  = \varepsilon^2s\pi\Big(c_7 s^{\frac{1}{k+r_0}}\Big)
  = \varepsilon^2\ell_\gamma(2s).
\end{align}
We now show that the homogeneous controller $u=\alpha(x)$ asymptotically stabilizes the auxiliary system (\ref{au_sys}) if we choose $\kappa\geq1$. The derivative of $V(x)$ along the closed-loop system (\ref{au_sys}) with $u=\alpha(x)$ is
\begin{align}\label{dv_as_ausys}
  \dot{V} & = L_fV + \ell_\gamma(2|L_{G_2}V|) + \frac{\kappa}{2}L_{G_1}V\alpha_s(x)
  \nonumber \\
  & = -M(x) - M_1(x) + |L_{G_2}V|\pi\Big(c_7|L_{G_2}V|^{\frac{1}{k+r_0}}\Big)
  -|L_{G_2}V|\pi\Big(\Gamma(x)\Big)
  \nonumber \\
  & \;\;\;\; \textup{[by (\ref{lgamma})]}
  \nonumber \\
  & \leq -M(x), \;\; \textup{[by (\ref{c1Ixc2})]}
\end{align}
where $M(x)$ is a function satisfying $M(0)=0$ and $M(x)=\frac{1}{2}\big[-\phi+\sqrt{\phi^2 + [L_{G_1}V(L_{G_1}V)^T]^2}\big]$ for all $x\in\R^n\backslash\{0\}$, and $M_1(x)$ is a function satisfying $M_1(0)=0$ and $M_1(x)=\frac{\kappa-1}{2}\big[\phi+\sqrt{\phi^2 + [L_{G_1}V(L_{G_1}V)^T]^2}\big]+\frac{\kappa}{2}c_{10} L_{G_1}V(L_{G_1}V)^T$ for all $x\in\R^n\backslash\{0\}$. With the homogeneity of $\phi(x)$ and $L_{G_1}V(x)$, we obtain that $M(x)$ and $M_1(x)$ are homogeneous of degree $2(k+r_0)>0$. Hence, $M(x)$ and $M_1(x)$ are continuous on the whole domain $\R^n$. Moreover, due to (\ref{LG2Vleadtophizero}), $M(x)$ is positive definite. Hence, auxiliary system (\ref{au_sys}) is asymptotically stabilized by the homogeneous controller $u=\alpha(x)$.

It follows from Theorem 1 that $u=2\alpha(x)=\kappa\alpha_s(x)$ is an inverse optimal controller of system  (\ref{sys}) with respect to the cost functional (\ref{Ju_de}) with $\beta=2$, $\lambda=2$, $E(x(t))=4 V(x(t))$, $R_1(x)=\vartheta^2 R(x)/\kappa $, $R_2(x)=R(x)/\kappa$, $\gamma_0(|w|)=4\ell_\eta(|w|/2)$, and
\begin{align}\label{lx_th2}
  l(x)
   & = 4\Big[M(x) + M_1(x) - |L_{G_2}V|\pi\Big(c_7|L_{G_2}V|^{\frac{1}{k+r_0}}\Big)
  \nonumber \\
  & \quad
  +|L_{G_2}V|\pi\Big(\Gamma(x)\Big)\Big]
  -h(x)^TR(x)h(x)/\kappa.
\end{align}
In (\ref{lx_th2}), $h(x)^TR(x)h(x)$ is continuous on $\R^n$. Because $h(x)$ is continuous on $\R^n$ and $R(x)$ is continuous on $\R^n\backslash\{0\}$, $h(x)^TR(x)h(x)$ is continuous on $\R^n\backslash\{0\}$.  Moreover, since $h(x)^TR(x)h(x)$ is a homogeneous function with degree $2(k+r_0)>0$, it follows from Lemma \ref{lem:bhat-homo-cont} that $h(x)^TR(x)h(x)$ is continuous on the whole domain $\R^n$. By Lemma \ref{lem:func-H}, one can choose a $\kappa$ such that $l(x)$ is positive definite. Since $l(x)$ is homogenous and positive definite, $l(x)$ is radially unbounded. Thus, the proof is completed. \hfill $\blacksquare$

\begin{remark}\rm
The HJI equation associated with the cost functional (\ref{Ju_de}) is
\begin{align}
    &L_fV + l(x) - \frac{1}{8}L_{G_1}VR_1(x)^{-1}(L_{G_1}V)^T
        \nonumber\\
    & \quad\quad\quad\quad
    + h(x)^TR_2(x)h(x) + \ell\gamma_0(|L_{G_2}V|) = 0.
    \label{eq:HJB-our-cost}
\end{align}
In our results, $l(x)$, $R_1(x)$ and $R_2(x)$ are constructed using $V(x)$. This is different from \cite{van_der_schaft_tac_1992, Isidori_1992, Isidori_1998}, where the weight matrices $R_1(x)$ and $R_2(x)$ are predefined by designers. As mentioned above, the continuous Lyaupnov function $V(x)$ is the value function of the cost functional (\ref{Ju_de}). Hence, this continuous $V(x)$ solves the HJI equation (\ref{eq:HJB-our-cost}).
\end{remark}

\begin{remark}\rm
Different from the HJI equation of \cite{krstic1998inverse}, our HJI equation (\ref{eq:HJB-our-cost}) additionally contains the term $h(x)^TR_2(x)h(x)$, which is related to the output penalty. We note that the continuity of $L_{G_1}VR_1(x)^{-1}(L_{G_1}V)^T$ and $h(x)^TR_2(x)h(x)$ is important for ensuring that the HJI equation (\ref{eq:HJB-our-cost}) has a continuous solution. This requirement makes the construction of our $R(x)$ in (\ref{optimal_u}) more challenging than that of \cite{krstic1998inverse}. In \cite{krstic1998inverse}, the weight matrix is constructed to be continuous on the boundary of the set $\{x\in\R^n:L_{G_1}V(x)=0\}$. Such a construction of $R(x)$ can guarantee that $L_{G_1}VR_1(x)^{-1}(L_{G_1}V)^T$ is continuous on $\R^n\backslash\{0\}$; however, it cannot guarantee the continuity of $h(x)^TR_2(x)h(x)$. To address this issue, we propose another construction for $R(x)$, which is continuous on $\R^n\backslash\{0\}$. This construction is inspired by \cite[Chapter 3.5.3]{Sepulchre_1997}.
\end{remark}

\begin{remark}\rm
If system (\ref{sys}) is homogeneously stabilizable in the case of $w\equiv0$, then the inverse optimal gain assignment problem for a nonlinear system that can be approximated by (\ref{sys}) is solvable (locally). Specifically, consider system
\begin{align}\label{sys_X}
    \dot{x} = \hat{f}(x) + \hat{G}_1(x)u + \hat{G}_2(x)w,
    \quad
    y = \hat{h}(x) + \hat{d}(x)u
\end{align}
where $\hat{f}(x) = f(x) + \bar{f}(x)$, $\hat{G}_1(x) = G_1(x) + \bar{G}_1(x)$, $\hat{G}_2(x) = G_2(x) + \bar{G}_2(x)$, $\hat{h}(x) = h(x) + \bar{h}(x)$ and $\hat{d}(x) = d + \bar{d}(x)$. Suppose that system (\ref{sys_X}) can be approximated by homogeneous system (\ref{sys}), that is, the functions $\bar{f}(x)$, $\bar{h}(x)$ and $\bar{d}(x)$ satisfy $\lim_{\varepsilon\rightarrow0}
\bar{f}(\Delta_\varepsilon^rx)/\varepsilon^k=0$,
$\lim_{\varepsilon\rightarrow0}
\bar{h}(\Delta_\varepsilon^rx)/\varepsilon^{k+r_0}=0$ and
$\lim_{\varepsilon\rightarrow0}
\bar{d}(\Delta_\varepsilon^rx)=0$ for $x\in\R^n\backslash\{0\}$, and the vector field $\bar{G}_j(x)$ for $j=1,2$, satisfies $\lim_{\varepsilon\rightarrow0}
\bar{G}_{ji}(\Delta_\varepsilon^rx)/\varepsilon^{\varsigma+r_i}=0$ for $x\in\R^n\backslash\{0\}$. By Theorem 2, there exist the functions $\gamma$ and $R(x)$, constant $\kappa_0$ and controller $u=\alpha(x)$ defined as in Theorem 1.  Then for $\kappa>\kappa_0$, the controller $u = \beta\hat{\alpha}(x)$, with
$\hat{\alpha}(x)  = -\frac{2\kappa}{\hat{d}(x)^T\hat{d}(x)}
  R(x)^{-1}(L_{\hat{G}_1}V)^T$,
solves the inverse optimal gain assignment problem (locally) of (\ref{sys_X}) by minimizing the cost functional (12) with $E(x)=2\beta V(x(t))$, $R_1=\hat{d}^TR\hat{d}/\hat{\kappa}$, $R_2=R/\hat{\kappa}$, $\gamma_0(|w|)=2\beta\lambda\gamma(\frac{|w|}{\lambda})$ and $l(x)=2\beta
\hat{H}_\kappa/\hat{\kappa}
+\beta(2-\lambda)[\ell_\gamma(2|L_{G_2}V|)+\ell_\gamma(2|L_{\bar{G}_2}V|)]
-\beta(\beta-2)L_{\hat{G}_1}V\hat{\alpha}(x)$, where $\hat{H}_\kappa=-\hat{\kappa}(L_{\hat{\tilde{f}}}V+L_{\hat{G}_1}V\hat{\alpha}(x))-\hat{h}^TR\hat{h}/(2\beta)$,  $\hat{\tilde{f}}= \tilde{f} + \bar{\tilde{f}}$ and $\bar{\tilde{f}}$ is of the form (\ref{f_tilde}) by replacing $f$ and $G_2$ with $\bar{f}$ and $\bar{G}_2$. Since $\hat{H}_\kappa\geq 4H_\kappa + o(\Gamma(x)^{2(k+r_0)})\geq0$ holds near the origin, $J(u)$ is meaningful and $J(u)_{\min}=2\beta V(x(0))$.
\end{remark}

Before the end of this section, we use an example to demonstrate the procedure of the inverse optimal design proposed in Theorem 2.

\begin{example}\rm
Consider the system
\begin{align}\label{ex_sys}
  \dot{x}_1  = -x_1 + x_2^3,\;\;
  \dot{x}_2  = u + w,\;\;
  y = [x_2,u]^T
\end{align}
which is homogeneous of degree 0 with respect to the dilation $(3,1)$. Herein $f(x)=[-x_1+x_2^3,0]^T$, $G_1(x)=G_2(x)=[0,1]^T$,  $h(x)=[x_2,0]^T$ and $\vartheta=1$ with $k=0$, $r_1=3$, $r_2=1$, $r_0=1$, $\varsigma=-1$. Choose a homogeneous Lyapunov function candidate with degree of $k+2r_0=2$ as $V(x) = (x_1^{4/3} + x_2^4)^{1/2}$. For any $x\in\R^n\backslash\{0\}$, we can compute
$L_fV(x)=\frac{2}{3}x_1^{1/3}
  (x_1^{4/3}+x_2^4)^{-1/2}
  (-x_1 + x_2^3)$,
$L_{G_1}V(x)= 2x_2^3(x_1^{4/3}+x_2^4)^{-1/2}$
and $L_{G_2}V(x)= 2x_2^3(x_1^{4/3}+x_2^4)^{-1/2}$.
Choose $\nu=4$ and $\Gamma(x) = (|x_1|^{4/3}+|x_2|^4)^{1/4}$. Take $\pi(s)=s$.
With (\ref{phi}), we have $\phi(x)
    = \frac{2}{3}x_1^{1/3}
  (x_1^{4/3}+x_2^4)^{-1/2}
  (-x_1 + x_2^3)
  + 2|x_2|^3(x_1^{4/3}+x_2^4)^{-1/4}
  $ for all $x\in\R^2\backslash\{0\}$.
Clearly, $\lim_{x\rightarrow0}\phi(x)=0$. For $x\in \R^2\backslash\{0\}$, $L_{G_1}V=0$ implies $x_2=0$. Thus, with (\ref{phi}),
\begin{equation}\label{ex_phi2}
  L_{G_1}V = 0 \;
  \Rightarrow
  \;
  \phi= \frac{2}{3}x_1^{\frac{1}{3}}
  \Big(x_1^{\frac{4}{3}}+0\Big)^{-\frac{1}{2}}(-x_1 + 0) + 0
  = - \frac{2}{3}x_1^{\frac{2}{3}}<0.
\end{equation}
Choose $\gamma(s)=\frac{1}{c_7}s^2$ and we have $\ell_\gamma(2s)=c_7s^2$, where $c_7=\frac{1}{2}$. With (\ref{au_sys}), the auxiliary system for (\ref{ex_sys}) is constructed as
\begin{align}\label{ex_au_sys}
  \dot{x}=\tilde{f}(x)+G_1(x)u
\end{align}
where $\tilde{f}(x)$ is a function satisfying $\tilde{f}(0)=f(0)=0$, and for any $x\neq0$, $\tilde{f}(x)=[-x_1 + x_2^3, x_2^3\big(x_1^{{4}/{3}}+x_2^4\big)^{-1/2}]^T$.
Design
\begin{align}\label{ex_u}
  u = \alpha(x) = \left\{
  \begin{array}{rll}
  & -\frac{\kappa}{2}R(x)^{-1}(L_{G_1}V)^T,  & x \neq 0 \\
  & 0, & x = 0
    \end{array} \right.
\end{align}
where $\kappa>1$ will be given latter and
\begin{equation}\label{ex_Rx}
  R(x) = \left\{
  \begin{array}{rll}
    & \Bigg[1+\frac{\Big(\phi + \sqrt{\phi^2 + [L_{G_1}V(L_{G_1}V)^T]^2}\Big)}{ L_{G_1}V(L_{G_1}V)^T}\Bigg]^{-1}, & L_{G_1}V \neq 0 \\
    & 1, & L_{G_1}V = 0
    \end{array} \right.
\end{equation}
From (\ref{ex_phi2}), (\ref{ex_u})-(\ref{ex_Rx}), $\alpha(x)$ is continuous on $\R^2 \backslash \{0\}$. Substituting (\ref{ex_u}) into (\ref{ex_au_sys}), we have
\begin{align}\label{ex_dv}
  \dot{V}(x)
  & = -M(x)-M_1(x)+ |L_{G_2}V|
  \nonumber \\
  & \quad \times
  \bigg[\Big(|x_1|^{\frac{4}{3}} + |x_2|^4\Big)^{-\frac{1}{2}}\Big[|x_2|^3-\Big(|x_1|^{\frac{4}{3}} + |x_2|^4\Big)^{\frac{3}{4}}\Big] \bigg]
\end{align}
where $M(x)$ is a function such that $M(0)=0$ and for any $x\neq0$, $M(x)=\frac{1}{2}[-\phi + \sqrt{\phi^2 + [L_{G_1}V(L_{G_1}V)^T]^2}]$; $M_1(x)$ is a function such that $M_1(0)=0$ and for any $x\neq0$, $M_1(x)=\frac{\kappa}{2} L_{G_1}V(L_{G_1}V)^T+\frac{\kappa-1}{2}[\phi+\sqrt{\phi^2 + [L_{G_1}V(L_{G_1}V)^T]^2}]$. Since $|x_2|^3-(|x_1|^{4/3} + |x_2|^4)^{3/4}\leq0$, it follows from (\ref{ex_phi2}) and (\ref{ex_dv}) that $\dot{V}(x)$ is negative definite. Now we compute $\kappa$ as in Lemma \ref{lem:func-H}. Let $Q=\{x\;|\; |x_1|\geq4|x_2|^3\}$. After computation, we have $\rho_1=0.66$, $\rho_2=0.24$, $\kappa_c=0.36$. $\rho=2.18$, $\rho_4=0.37$, $\rho_3=0.42$ and $\kappa_1=10.55$. Choose $\kappa=11$ and $\beta=\lambda=2$, then $u=\alpha^*(x)=2\alpha(x)$ is an inverse optimal controller of (\ref{ex_sys}) with respect to (\ref{Ju_de}) with $E(x)=4V(x)$, $R_1(x)=R_2(x)=R(x)/11$ in (\ref{ex_Rx}), $\gamma_0(|w|)=2w^2$, and $l(x)$ is a function satisfying $l(0)=0$ and for any $x\neq0$, $l(x)
   = 4\{M(x) + M_1(x)
   + 2|x_2|^3(|x_1|^{4/3} + |x_2|^4)^{-1}
  [(|x_1|^{4/3} + |x_2|^4)^{3/4}-|x_2|^3]\}
  -x_2^2R(x)/11$.
\end{example}

\section{Conclusion}

In this work, we pose and solve a new inverse optimal gain assignment problem, where the cost functional puts penalties on the output, in addition to the state, control input and disturbance as in current works. The benefit of penalizing the output is that the resulting inverse optimal controllers can ensure both ISS and IOS. Although penalizing the output provides an IOS guarantee, it is difficult to construct the corresponding cost functionals, which must be proved meaningful. We use homogeneity properties to overcome this challenge and provide sufficient conditions for solving the inverse optimal gain assignment problem. We show that homogeneous stabilizability of the unperturbed homogeneous systems is sufficient for the solvability of inverse optimal gain assignment problem for homogeneous systems.


\begin{thebibliography}{10}



\bibitem{Kawski1990ctat}
Kawski, M. (1990).
\newblock Homogeneous stabilizing feedback laws.
\newblock {\em Control Theory and Advanced Technology}, 6(4): 497-516.

\bibitem{hong_2001_tac}
Hong, Y., Huang, J., \& Xu, Y. (2001).
\newblock On an output feedback finite-time stabilization problem. \newblock {\em IEEE Transactions on Automatic Control}, 46(2): 305-309.

\bibitem{Hermes_1991_de}
Hermes, H. (1991).
\newblock Homogeneous coordinates and continuous asymptotically stabilizing feedback control.
\newblock {\em Differential Equations, Stability and Control}, 249.

\bibitem{Rosier1992scl}
Rosier, L. (1992).
\newblock Homogeneous Lyapunov function for homogeneous continuous vector field.
\newblock {\em Systems} \& {\em Control Letters}, 19(6): 467-473.

\bibitem{Polyakov_2023}
Polyakov, A., \& Krstic, M. (2023).
\newblock Finite- and fixed-time nonovershooting stabilizers and safety filters by homogeneous feedback.
\newblock{\em IEEE Transactions on Automatic Control}, 68(11): 6434-6449.

\bibitem{hong2001auto_homogenous}
Hong, Y. (2001).
\newblock $H_\infty$ control, stabilization, and input-output stability of nonlinear systems with homogeneous properties.
\newblock {\em Automatica}, 37(6): 819-829.

\bibitem{Nakamura2009tac}
Nakamura, N., Nakamura, H., Yamashita, Y., \& Nishitani, H. (2009). \newblock Homogeneous stabilization for input affine homogeneous systems.
\newblock {\em IEEE Transactions on Automatic Control}, 2009, 54(9): 2271-2275.

\bibitem{Moulay2008tac}
Moulay, E. (2008).
\newblock Stabilization via homogeneous feedback controls.
\newblock {\em Automatica}, 2008, 44(11): 2981-2984.


\bibitem{kalman1964linear}
Kalman, R. E. (1964).
\newblock When is a linear control system optimal?
\newblock {\em Transactions of the ASME, Series D, Journal of Basic Engineering}, 86, 51--60.

\bibitem{freeman1996inverse}
Freeman, R. A., \& Kokotovic, P. V. (1996).
\newblock Inverse optimality in robust stabilization.
\newblock {\em SIAM Journal on Control and Optimization}, 34(4), 1365--1391.

\bibitem{krstic1998inverse}
Krstic, M., \& Li, Z. H. (1998).
\newblock Inverse optimal design of input-to-state stabilizing nonlinear controllers.
\newblock {\em IEEE Transactions on Automatic Control}, 43(3): 336--350.

\bibitem{deng1997stochastic}
Deng, H., \& Krstic, M. (1997).
\newblock Stochastic nonlinear stabilization-ii: inverse optimality.
\newblock {\em Systems} \& {\em Control Letters}, 32(3): 151--159.

\bibitem{li1997optimal}
Li, Z. H., \& Krstic, M. (1997).
\newblock Optimal design of adaptive tracking controllers for nonlinear systems.
\newblock {\em Automatica}, 33(8): 1459--1473.

\bibitem{lu2023auto}
Lu, K., Liu, Z., Yu, H., Chen, C. P., \& Zhang, Y. (2024).
\newblock A small-gain approach to inverse optimal adaptive control of nonlinear systems with unmodeled dynamics.
\newblock {\em Automatica}, 2024(159): 111360.



\bibitem{Isidori_1998}
Isidori, A., \& Lin, W. (1998).
\newblock{Global $L_2$-gain design for a class of nonlinear systems}.
\newblock{\em Systems} \& \newblock{\em Control Letters}, 34(5):295-302.


\bibitem{Sontag1995scl}
Sontag, E. D., \& Yuan, W. (1995).
\newblock On characterizations of the input-to-state stability property.
\newblock {\em Systems} \& {\em Control Letters}, 24(5): 351-359.



\bibitem{IOS_book_2009}
Desoer, C. A., \& Vidyasagar, M. (2009).
\newblock{Feedback systems: input-output properties}.
\newblock{\em Society for Industrial and Applied Mathematics}.

\bibitem{Doyle1989tac}
Doyle, J. C., Glover, K., Khargonekar, P. P., \& Francis, B. A. (1989).
\newblock State-space solutions to standard $H_2$ and $H_\infty$ control problems.
\newblock {\em IEEE Transactions on Automatic Control}, 34(8): 831-847.

\bibitem{Bhat_mcss_2005}
Bhat, S. P., \& Bernstein, D. S. (2005).
\newblock Geometric homogeneity with applications to finite-time stability.
\newblock {\em Mathematics of Control, Signals and Systems}, 17: 101-127.



\bibitem{Sontag_scl_1989}
Sontag, E. D. (1989).
\newblock A universal construction of Artstein's theorem on nonlinear stabilization.
\newblock {\em Systems} \& {\em Control Letters}, 13(2): 117-123.

\bibitem{Isidori_1992}
Isidori, A., \& Astolfi, A. (1992).
\newblock Disturbance attenuation and $H_\infty$-control via measurement feedback in nonlinear systems.
\newblock {\em IEEE Transactions on Automatic Control}, 37(9): 1283-1293.

\bibitem{Belhadjoudja_2024_ECC}
Belhadjoudja, M. C., Krstic, M., Maghenem, M. \& Witrant, E. (2024).
\newblock Inverse optimal Cardano-Lyapunov feedback for PDEs with convection.
\newblock{\em European Control Conference (ECC)}, 3823-3828.


\bibitem{Krstic_inverse_safety_2024}
Krstic, M. (2023).
\newblock Inverse optimal safety filters.
\newblock{\em IEEE Transactions on Automatic Control}, 69(1): 16-31.

\bibitem{Ryan_1995}
Ryan, E. P. (1995).
\newblock Universal stabilization of a class of nonlinear systems with homogeneous vector fields.
\newblock {\em Systems} \& {\em Control Letters}, 26(3): 177-184.

\bibitem{Andrieu_V_2008}
Andrieu, V., Praly, L., \& Astolfi, A. (2008).
\newblock Homogeneous approximation, recursive observer design, and output feedback.
\newblock{\em SIAM Journal on Control and Optimization}, 47(4): 1814-1850.

\bibitem{Bernuau_2013}
Bernuau, E., Polyakov, A., Efimov, D., \& Perruquetti, W. (2013).
\newblock Verification of ISS, iISS and IOSS properties applying weighted homogeneity.
\newblock {\em Systems} \& {\em Control Letters}, 62(12): 1159-1167.

\bibitem{Bernuau_2014}
Bernuau, E., Efimov, D., \& Perruquetti, W. (2014).
\newblock On the robustness of homogeneous systems and a homogeneous small gain theorem.
\newblock{\em IEEE Conference on Control Applications (CCA)}, 929-934.

\bibitem{qian_2015}
Qian, C., Lin, W., \& Zha, W. (2015).
\newblock{Generalized homogeneous systems with applications to nonlinear control: a survey}.
\newblock{\em Math.Control and Related Fields}, 5(3): 585-611.

\bibitem{zhao_2022}
Zhao, C., \& Lin, W. (2022).
\newblock{Homogeneity, forward completeness, and global stabilization of a family of time-delay nonlinear systems by memoryless non-Lipschitz continuous feedback}.
\newblock{\em IEEE Transactions on Automatic Control}, 67(11): 5916-5931.

\bibitem{yu_2023}
Yu, X., \& Lin, W. (2023).
\newblock{Global input delay tolerance of MIMO nonlinear systems under nonsmooth feedback: a homogeneous perspective}.
\newblock{\em IEEE Transactions on Automatic Control}, 68(7): 3992-4007.

\bibitem{coron_MPC}
Coron, J. M., Grunne, L., \& Worthmann, K. (2020).
\newblock Model predictive control, cost controllability, and homogeneity.
\newblock{\em SIAM Journal on Control and Optimization}, 58(5): 2979-2996.

\bibitem{Polyakov_2020}
Polyakov, A. (2020).
\newblock{Generalized homogeneity in systems and control}.
\newblock{\em Cham: Springer International Publishing}.

\bibitem{van_der_schaft_tac_1992}
van der Schaft, A. J. (1992).
\newblock{$L_2$-gain analysis of nonlinear systems and nonlinear state feedback $H_\infty$ control}.
\newblock{\em IEEE Transactions on Automatic Control}, 37(6): 770-784.

\bibitem{krstic_ijc_1994}
Krstic, M., \& Kokotovic, P. V. (1994).
\newblock{Observer-based schemes for adaptive nonlinear state-feedback control}.
\newblock{\em International Journal of Control}, 59(6): 1373-1381.

\bibitem{Krstic_1995}
Krstic, M., Kanellakopoulos, I., \& Kokotovic, P. V. (1995).
\newblock{Nonlinear and adaptive control design}.
\newblock{\em New York: Wiley}.

\bibitem{sontag-integral-ISS}
Sontag, E. D. (1998).
\newblock{Comments on integral variants of ISS}.
\newblock {\em Systems} \& {\em Control Letters}, 34(1-2): 93-100.

\bibitem{Anderson_1990}
Anderson, B. D. O., \& Moore, J. B. (1990).
\newblock{Linear optimal control}.
\newblock{\em New Jersey: Prentice Hall}.


\bibitem{Moylan_1973_tac}
Moylan, P., \& Anderson, B. D. O. (1973).
\newblock{Nonlinear regulator theory and an inverse optimal control problem}.
\newblock {\em IEEE Transactions on Automatic Control}, 18(5): 460-465.

\bibitem{Sepulchre_1997}
Sepulchre, R., Jankovic, M., \& Kokotovic, P. V. (1997).
\newblock{Constructive nonlinear control}.
\newblock{\em London: Springer-Verlag}.



\bibitem{Linwei_ijc_1996}
Lin, W. (1996).
\newblock{Mixed $H_2/H_\infty$ control via state feedback for nonlinear systems}.
\newblock {\em International Journal of Control}, 64(5): 899-922.



\end{thebibliography}

\end{document}